\documentclass[aps,twocolumn,showpacs,preprintnumbers,amsmath
,nofootinbib
,superscriptaddress,showkeys,prd]{revtex4}

\usepackage{epsfig}
\usepackage{graphicx}
\usepackage{mathptmx}      
\usepackage{latexsym}
\usepackage{amsmath}
\usepackage{amsfonts}
\usepackage{bm}

\def\slashchar#1{\setbox0=\hbox{$#1$}
   \dimen0=\wd0 \setbox1=\hbox{/} \dimen1=\wd1
   \ifdim\dimen0>\dimen1 \rlap{\hbox to \dimen0{\hfil/\hfil}} #1
   \else  \rlap{\hbox to \dimen1{\hfil$#1$\hfil}} / \fi}

\def\tr{{\rm tr}}

\newcommand{\fm}{\,\mathrm{fm}}
\newcommand{\MeV}{\,\mathrm{MeV}}
\newcommand{\GeV}{\,\mathrm{GeV}}
\newcommand{\SU}{{\rm SU}}

\newcommand{\Eq}[1]{Eq.~(\ref{eq:#1})}

\newcommand{\vx}{{\bm{x}}}
\newcommand{\vp}{{\bm{p}}}

\newcommand{\vP}{{\bm{P}}}
\newcommand{\ignore}[1]{}

\newcommand{\cL}{{\mathcal L}}
\newcommand{\cO}{{\mathcal O}}
\newcommand{\Or}{{O}}
\newcommand{\DF}{{F}}
\newcommand{\DS}{{S}}
\newcommand{\DU}{{U}}
\newcommand{\bQ}{\bar{Q}}
\newcommand{\bq}{\bar{q}}

\newcommand{\bM}{\bar{M}}
\newcommand{\bH}{\bar{H}}

\begin{document}

\title{Heavy quark-antiquark free energy and thermodynamics of string-hadron
  avoided crossings}

\author{E. Meg\'{\i}as}
\email{emegias@mppmu.mpg.de}

\affiliation{Max-Planck-Institut f\"ur Physik (Werner-Heisenberg-Institut), F\"ohringer Ring 6, D-80805 Munich, Germany}

\author{E. \surname{Ruiz Arriola}}
\email{earriola@ugr.es}

\author{L. L. Salcedo}
\email{salcedo@ugr.es}

\affiliation{Departamento de F\'{\i}sica At\'omica, Molecular y Nuclear and
  Instituto Carlos I de F\'{\i}sica Te\'orica y Computacional \\ Universidad
  de Granada, E-18071 Granada, Spain.}

\date{\today}

\begin{abstract}
\medskip
The correlation function between two Polyakov loops encodes the free-energy
shift due to a pair of separated colour conjugated sources in the hot QCD
medium. This is analyzed in terms of a novel K\"all\'en-Lehmann spectral
representation for the separating distance, implying an increasing and concave
free-energy at all temperatures.  We express the heavy $\bQ Q$ free-energy
shift below the phase transition in QCD in terms of colour neutral purely
hadronic states with no explicit reference to quarks and gluons. Good
agreement with lattice data is achieved when considering the avoided crossing
mechanism underlying string breaking and with standard quenched values of the
string tension known from charmonium and bottomonium phenomenology.  We also
address the role of the corresponding entropy shift and its renormalization
group properties.
\end{abstract}

\pacs{11.10.Wx 11.15.-q  11.10.Jj 12.38.Lg }

\keywords{finite temperature; heavy quarks;
free-energy; Polyakov Loop}

\maketitle

\section{Introduction}
\label{sec:Introduction}

Long before the advent of QCD the thermodynamic and hydrodynamic
interpretation of high energy collisions in terms of hadrons was envisaged
after the pioneering works of Fermi~\cite{Fermi:1950jd} and
Landau~\cite{Landau:1953gs} (for a review see
e.g.~\cite{Florkowski:2010zz}). A major inspiring breakthrough came about when
Hagedorn found that purely hadronic matter forming a hadron resonance gas
(HRG) has an exponentially growing level density implying the existence of a
hadronic limiting temperature~\cite{Hagedorn:1965st} of about $T_H =150 \MeV$
(see \cite{Broniowski:2004yh} for an upgrade and \cite{Rafelski:2016hnq} for a
historical overview). According to the quantum virial expansion this
corresponds to a weakly interacting multicomponent gas of
resonances~\cite{Dashen:1969ep}. The discovery of asymptotic freedom and QCD
in terms of quarks and gluons carrying experimentally elusive colour degrees
of freedom triggered an intensive research on the phase structure over the
last half a century and has motivated a dedicated experimental effort at SPS,
RHIC and LHC facilities within the heavy ions collisions traveling at almost
the speed of light~\cite{Florkowski:2010zz}. Lattice calculations provide
direct access to the equation of state (EoS) and a strong evidence that a
crossover between a hadron gas and a quark-gluon liquid phase takes place at a
temperature of about $150 \MeV$~\cite{Aoki:2006we}. The final consensus on the
EoS has only been achieved recently buttressing in passing the venerable HRG
in the subcritical region (see e.g.~\cite{Fodor:2012rma,Petreczky:2012rq} for
reviews) and promoting this model as a canonical comparison tool for lattice
calculations monitoring non-hadronic
effects~\cite{Huovinen:2009yb,Ding:2015ona}. Besides, modern hydrodynamical
descriptions of ultrarelativistic heavy ions provide in a unified way the
formation of the fireball as well as the subsequent evolution process just in
terms of the EoS in the gas-liquid crossover interregnum
\cite{Florkowski:2010zz}.

In a non-abelian gauge theory such as QCD one can meaningfully address
specific information on quark and gluon colour charges and their interactions
in the hot medium beyond the EoS which could not be posed in the pre-QCD
times. Nonetheless, we expect that any physical observable can still be described
in terms of manifestly colour singlet hadronic states below the phase
transition.  We have recently shown this quark-hadron duality at finite
temperature for the quark self-energy via the renormalized Polyakov
loop~\cite{Megias:2012kb} using single heavy meson charm and bottom excited
spectra (see also \cite{Bazavov:2013yv}). (See ~\cite{Megias:2013xaa} for
higher representations and \cite{Arriola:2014bfa} for a pedagogical
discussion.) In this letter we verify that the quark-antiquark free-energy can
be represented, as naively expected, via purely hadronic states and a single
bosonic string potential, with no explicit reference to quarks and
gluons. This requires a non-trivial level avoided crossing structure in the
heavy-light meson-antimeson spectrum and is done in harmony with a new
Laplace-Stietjels spectral representation of the Polyakov correlator, whence
concavity properties of the free-energy at any temperature can be trivially
deduced.  Some motivation for the present work has been presented previously
in \cite{Arriola:2015gra,Megias:2015qya}.

The quark-antiquark free energy is ambiguous by a constant which can be fixed
at a reference temperature and relative distance between the heavy sources. On
the contrary both the entropy shift and the specific heat shift are
unambiguous, and one expects them to fulfill renormalization group
invariance. In a short note we have profited from this view to display the
entropy shift due to a single heavy source based on integrating the specific
heat with suitable boundary conditions \cite{Megias:2016bhk}. In the present
work we extend such an analysis to the corresponding quark-antiquark situation
and show that more information than the specific heat is needed in order to
reconstruct the free energy.

The paper is structured as follows. In
Section~\ref{sec:SpectralRepresentation} we present a derivation of the
K\"allen-Lehman representation for the free energy of two colour charged
conjugated sources. This allows to easily deduce convexity properties for the
free energy. In Section~\ref{sec:renormalization_group} we analyze the
renormalization group aspects of free energy, entropy and specific heat and
illustrate the constraints for the case of a medium and the modifications
following the addition of a colour source and its charge conjugated source at
a given separation distance. In Section \ref{sec:confined_phase} we analyze
the avoided crossing structure of the spectrum made of a string and pairs of
heavy-light, $\bq Q$ and singly heavy baryons $ Q qq $ and the corresponding
partition function containing the Hadron Resonance Gas and a string. We
highlight the important role played by the string-hadron transition on the
light of a comparison with available recent lattice data for the Polyakov loop
correlator. Finally in Section~\ref{sec:Conclusions} we summarize our points
and come to the conclusions.

\section{K\"all\'en-Lehmann spectral representation for the correlation of two Polyakov loops}
\label{sec:SpectralRepresentation}

When two heavy colour charge conjugated sources belonging to the
representations $R$ and $\bar R$ of the colour gauge group $\SU(N_c)$ are
created and placed at a given distance in the hot medium, there arises
a free-energy which provides the maximum work the system can exchange with the
medium at a fixed temperature. McLerran and Svetitsky suggested 35 years ago
to explore these free-energies for the fundamental representation as suitable
order parameters for a hadronic-quark gluon plasma phase
transition~\cite{McLerran:1980pk,McLerran:1981pb} (for an early review see
e.g.~\cite{Svetitsky:1985ye}).

\subsection{Standard scheme}
\label{subsec:standard}

\begin{figure}[t]
\begin{center}
\epsfig{figure=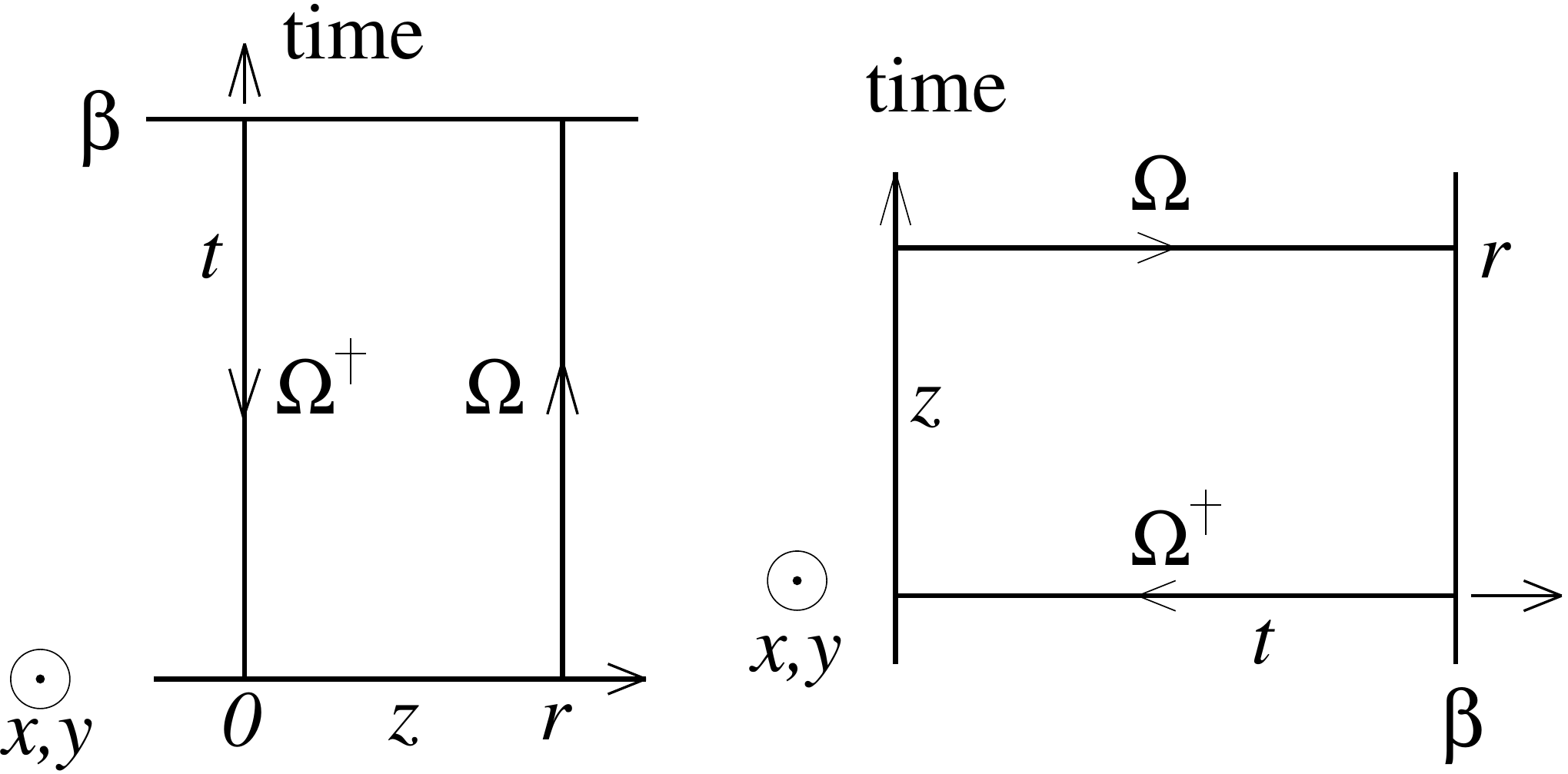,height=45mm,width=85mm}
\end{center}
\caption{Two schemes for $\langle\tr(\Omega(r))\tr(\Omega^\dagger(0))\rangle$.
In both cases the Hamiltonian evolves in the vertical direction.}
\label{fig:spec1}
\end{figure}

The obvious approach to the heavy quark-antiquark free-energy is to consider a
compactified Euclidean time to include the finite temperature, so the three
coordinates $x,y,z$ are spatial ones and $t$ is time-like (see
Fig.~\ref{fig:spec1}, left). Denoting $Z_0$ the partition function of the
system without sources, $Z_{R\otimes\bar{R}}$ the partition function with
heavy sources a distance $r$ apart, and $C$ the correlation function between
Polyakov loops,
\begin{equation}
C(r,T) = \langle \tr_R \Omega (r)
\tr_{\bar{R}} \Omega (0)\rangle_T
=  \frac{Z_{R\otimes\bar{R}}(r,T)}{Z_0(T)},
\label{eq:1}
\end{equation}
where the Polyakov loop, $\Omega = {\mathcal P} e^{i\int_0^{1/T} \! g A_0 dt}$, is a
purely gluonic operator, $\tr_R (1)= \mathrm{dim} \,R$, and $\langle ~
\rangle_T$ is the thermal expectation value. Correspondingly, for the shift in
the free-energy
\begin{equation}
\DF(r,T) \equiv F_{R \otimes \bar R}(r,T) - F_0(T) = 
-T \log C(r,T)
.
\end{equation}
Cluster decomposition and translational invariance requires factorization
(using also colour charge conjugation $\langle \tr_R \Omega\rangle_T = \langle
\tr_{\bar R} \Omega\rangle_T$)
\begin{equation}
C(\infty,T) = \langle \tr_R \Omega\rangle_T^2 = e^{-2\DF(T)/T}
\end{equation}
where $\DF(T)$ is the shift in the free-energy produced by introducing a
single heavy charge.

To avoid paradoxes it is important to note that, unlike $F_{R \otimes
  \bar{R}}(r,T)$ and $F_0(T)$, the shift $\DF(r,T)$ is not a true free-energy.
This quantity is not extensive and thermodynamic stability does not require
the entropy shift
\begin{equation}
\DS(r,T) = - \frac{ \partial \DF(r,T) }{\partial T}
,
\end{equation}
to be an increasing function of the temperature. Likewise, $C(r,T)$ is not a
true partition function (i.e., a sum of states with non negative integer
degeneracies) but rather the ratio of two partition
functions~\cite{Luscher:2002qv}. Nevertheless, by the same arguments given in
~\cite{Megias:2012kb,Megias:2013xaa} for the Polyakov loop, this ratio should
behave, to a good approximation, like a partition function for temperatures
below the crossover, that is,
\begin{equation} \begin{split}
C(r,T)  \approx \sum_n g_n e^{-E_n(r)/T} = \int_{-\infty}^\infty dE \rho (E,r) e^{-E/T} ,
\end{split}
\label{eq:4}
\end{equation}
where $\rho (E,r)$ is non negative. This implies that standard thermodynamic
relations apply to $F(r,T)$ in this low temperature regime. Similar remarks
apply for $\langle \tr_R \Omega \rangle$ and
$F(T)$~\cite{Megias:2012kb,Megias:2013xaa}.

\subsection{Alternative scheme}
\label{subsec:alternative}

Additional information can be obtained from the alternative point of view in
which $x,y,t$ are regarded as spatial coordinates and $z$ as Euclidean time
(see Fig.~\ref{fig:spec1}, right). In this picture $\beta=1/T$ (the inverse
physical temperature) is the size of the system in the spatial $t$
direction. The size remains infinite in the $x$ and $y$ directions. The system
is described by a Hamiltonian $H_z(T)$ which produces the evolution in the $z$
direction and depends on $T$. Since the size in the $z$ direction is unbounded
the effective temperature is zero. Therefore, in this picture, the expectation
values are taken in the vacuum state of $H_z(T)$, which we denote
$|0;T\rangle$.

To carry out a canonical treatment, a convenient choice is the static
gauge $\partial_t A_t=0$ (where $A_t$ is the gluon gauge field in the $t$
direction). This leaves the Polyakov loop in the $t$ direction, $\Omega(x,y)$,
together with $A_{x,y,z}$ and $q$, $\bq$, as degrees of freedom. $\Omega$ acts
in the Hilbert space as a multiplicative operator. In this view the
expectation value of two Polyakov loops, one at $x=y=z=0$ and the other at
$x=y=0$, and a later time $z=r$, represents the action of the operator ${\rm
  Tr}_R\Omega^\dagger$ (at $x,y=0$) on the vacuum $|0,T\rangle$ at time $z=0$
followed by evolution to time $z=r\,$ to connect again with ${\rm
  Tr}_R\Omega^\dagger|0,T\rangle$
\begin{equation}
C(r,T) = \langle 0,T | {\rm Tr}_R \Omega \, e^{-r H_z(T)} 
{\rm Tr}_R \Omega^\dagger  |0,T\rangle
.
\end{equation}
This allows to establish an exact spectral representation valid for any
representation $R$ and any temperature, by inserting a complete set of
eigenstates of $H_z$, $H_z|n,T\rangle=w_n(T)|n,T\rangle$,
\begin{equation}\begin{split}
C(r,T) &=
\sum_n |\langle n,T | {\rm Tr}_R \Omega^\dagger 
|0,T\rangle|^2  \, e^{-r w_n(T)}
\\
&:=
\int_0^\infty dw \, \tau_1(w,T) \, e^{-r w}
.
\label{eq:L2}
\end{split}\end{equation}
The density of states excited by the Polyakov loop, $\tau_1(w,T)$, is a
non-negative function with support on $ w \ge 0$,\footnote{Note that the
  vacuum has zero energy, since replacing ${\rm Tr}_R \Omega^\dagger$ by the
  identity operator would yield $C=1$ for all $r$.} so \Eq{L2} is a
Laplace-Stietjels transformation~\cite{book:1304498}.

The sum rule naturally splits into its $r$-independent disconnected component,
coming from the vacuum term (i.e., the term $n=0$ in the sum), and the
connected component from the excited states:
\begin{equation}
C(r,T) = L(T)^2 + C_c(r,T)
,\quad L(T) \equiv \langle \tr_R\Omega \rangle_T
.
\label{eq:L2a}
\end{equation}
The disconnected component yields a contribution $L(T)^2 \,\delta(w)$ in
$\tau_1(w,T)$.

In addition to the usual scheme $(x,y,z;t)$ and the one considered here,
$(x,y,t;z)$, a third scheme would be $(x,z,t;y)$ $y$ being now the time
direction. In this case $C(r,T)$ is just the expectation value of an operator,
${\rm Tr}_F \Omega(r){\rm Tr}_F \Omega^\dagger(0)$, acting at time $y=0$ on
the vacuum $|0,T\rangle$ (the same vacuum as in the $(x,z,t;y)$ scheme). No
excited states are involved in this case.

\subsection{Convexity properties of the free energy}
\label{subsec:convexity}

An interesting consequence of the spectral relation (\ref{eq:L2}) comes from
the fact that $\tau_1(w,T)$ is positive and has positive support. This
automatically implies that $C(r,T)$ is an absolutely monotonic function with
respect to $r$, that is, 
\begin{equation}
(-1)^n \frac{\partial^n C(r,T)}{\partial r^n} \ge 0
,\quad n=0,1,2,\ldots
\end{equation}
Further bounds on the free-energy can be derived by rewriting these
inequalities as expectation values with $\tau_1(w,T) e^{-r w}$ as  measure:
\begin{equation}
\frac{(-1)^n}{C}\frac{\partial^n C}{\partial r^n} 
=
\frac{1}{C}\int_0^\infty dw \, \tau_1 e^{-r w} w^n
\equiv 
\langle w^n \rangle
.
\end{equation}
In this  view $\langle w^n \rangle \ge 0$ because both $w$ and the
measure are non negative. For positive $r$, the measure is exponentially
convergent at large $w$, since that regime of energies is dominated by the
Coulomb interaction between the two conjugated charges (see below).

Letting $f \equiv \DF(r,T)/T = -\log C$, such inequalities imply, 
\begin{equation}
f^\prime = \langle w \rangle \ge 0,
\quad
f^\prime{}^2 - f^{\prime\prime}= \langle w^2 \rangle \ge 0
, \ldots
\end{equation}
The prime denotes derivative with respect to $r$.  Tighter bounds can be
obtained by using optimized positive polynomials.\footnote{The coefficients of
  the polynomials have been chosen so as to remove higher powers of $1/T$,
  i.e., monomials with many $f$'s. The coefficients in front of $f^\prime
  f^{\prime\prime}$ and $f^{\prime\prime}{}^2$ cannot be reduced since in that
  case the expressions would not be positive definite for $C(r,T)=1+a e^{-r}$
  and arbitrary positive $a$.} Specifically
\begin{equation}\begin{split}
\langle w \rangle &= f^\prime \ge 0,
\\
\langle (w-\langle w \rangle)^2 \rangle &=  - f^{\prime\prime} \ge 0
\\
\langle w(w-\langle w \rangle)^2 \rangle &=  f^{\prime\prime\prime}-f^\prime
f^{\prime\prime} \ge 0
\\
\langle (
w^2-2 \langle w \rangle w +2 \langle w \rangle^2 - \langle w^2 \rangle
)^2 \rangle
&=
2f^{\prime\prime}{}^2-f^{(iv)} \ge 0
.
\end{split}\end{equation}
These inequalities hold for all $(r,T)$. The first two relations imply
\begin{equation}
\frac{\partial \DF(r,T)}{\partial r} \ge 0 
,\qquad
\frac{\partial^2
  \DF(r,T)}{\partial r^2}  \le 0
.
\label{eq:8}
\end{equation}
An immediate consequence is that the zero temperature heavy quark-antiquark
potential $V(r)=\DF(r,0)$ must be an increasing and concave function of $r$,
\begin{equation}
V^\prime(r)\ge 0
,\qquad
V^{\prime\prime}(r)\le 0
,
\end{equation}
a result previously established from reflection
positivity~\cite{bachas1986concavity,Nussinov:2000kn}.  A similar
representation to ours on the lattice was proposed long
ago~\cite{Kiskis:1985vf} although the important new implications discussed
here were not addressed. A further property from the Laplace-Stietjels
representation is that the analytical continuation to the complex plane $r$
has no singularities for ${\rm Re}(r) > 0$.

\subsection{K\"all\'en-Lehmann spectral representation}
\label{subsec:KL}

The spectral relation in \Eq{L2} can be improved by taking into account full
translational and rotational invariances in the three non-compactified
directions. Indeed, up to now we have only considered expectation values of
$e^{-r H_z}$ in a single state namely, $\tr_R \Omega^\dagger(\mathbf{0})|0
\rangle$ (here $\mathbf{0}$ is the origin of the $(x,y)$ plane), however, one
can take more general states of the type $|\psi\rangle = \sum_i \psi_i \tr_R
\Omega^\dagger(\mathbf{r}_i)|0 \rangle$ and still $\langle \psi |e^{-z H_z}|
\psi \rangle$ must be positive for any choice of the coefficients
$\psi_i$. Equivalently, the function
\begin{equation}
h(\mathbf{r}_1,\mathbf{r}_2;z) \equiv \langle 0 | \tr_R \Omega(\mathbf{r}_1)
e^{-z H_z} \tr_R \Omega^\dagger(\mathbf{r}_2) | 0 \rangle ,\quad z\ge 0 ,
\end{equation}
provides the matrix elements of a certain positive operator $\cO(z)$
in $L^2(\mathbb{R}^2)$,
\begin{equation}
h(\mathbf{r}_1,\mathbf{r}_2;z) = \langle \mathbf{r}_1 | \cO(z) |
\mathbf{r}_1 \rangle ,
\end{equation}
and furthermore
\begin{equation}
h = h(|\mathbf{r}_1 - \mathbf{r}_2|^2+z^2)
.
\end{equation}

In view of the translational invariance, the simplest way to impose the
positivity condition on $\cO(z)$ is to work in momentum space where
this operator is diagonal. The details of the derivation are given in
Appendix \ref{app:AKL}. One finds a tighter exact spectral representation for
the connected component of $C(r,T)$
\begin{equation}
C_c(r,T) = \int_0^\infty d\mu \tau(\mu,T) \frac{e^{-\mu r}}{4\pi r}
,
\label{eq:n12}
\end{equation}
for some non-negative density $\tau(\mu,T)$.  This is nothing else than the
usual K\"all\'en-Lehmann spectral representation but in three dimensions:
$\tr_R\Omega(x,y,z)$ behaves like an ordinary scalar field at zero temperature
in a three dimensional space-time with Lorentz invariance. As it turns out, a
formula equivalent to the one derived here was already noted in
\cite{Luscher:2004ib} in the context of Yang-Mills theory with unbroken center
symmetry.

The new spectral relation contains the previous one, \Eq{L2}, through the relation
\begin{equation}
\tau_1(w,T)  = L(T)^2 \delta(w) + \frac{1}{4\pi} \int_0^w d\mu \tau(\mu,T)
.
\end{equation}
As a consequence we learn that the connected component of $\tau_1(w,T)$ is not
only positive but also an increasing function of $w$. Also it is important to
note that $w$ is an energy while $\mu$ is an invariant mass, thus the support
of $\tau(\mu,T)$ may have a discrete part below the two particle threshold.

When the Polyakov loop is not present, $L(T)=0$ (e.g., Yang-Mills in the
confined phase) it is easy to derive tighter conditions on the free energy
shift from the new representation in \Eq{n12}, since then $r\, C(r,T)$ has all
the properties previously exploited for $C(r,T)$. So in this case, again
letting $f=-\log(C)$, it follows that $f^\prime \ge 1/r$, $-f^{\prime\prime}
\ge 1/r^2$, $f^{\prime\prime\prime}-f^\prime f^{\prime\prime} \ge 3/r^3$,
etc. In particular,
\begin{equation}
\frac{\partial \DF}{\partial r} \ge \frac{T}{r}
,\quad
\frac{\partial^2 \DF}{\partial r^2} \le -\frac{T}{r^2}
\qquad (L(T)=0)
.
\end{equation}
The first relation implies that, at finite temperature, $\DF(r,T)$ must
increase at least logarithmically (namely, as $T\log(r)$) with the separation,
as a direct consequence of the factor $1/(4\pi r)$ in the sum rule, whenever
the Polyakov loop vanishes.

The determination of the invariant-mass spectral density $\tau(\mu,T)$ seems
of interest since this quantity contains much condensed information on the
system composed of two heavy conjugated charges at finite temperature. As it
turns out, such determination can be carried out for in some simple cases.
A first case is
\begin{equation}
F(r,T) = -\frac{\alpha}{r}
\end{equation}
which should describe quenched QED (a free theory) and entails $L(T)=1$ at all
temperatures.  Let
\begin{equation}
G_n(x) \equiv 
\frac{I_n(2\sqrt{x})}
{x^{n/2}} 
= 
\frac{{}_0F_1(n+1;x)}{n!}
=
\sum_{k=0}^\infty \frac{x^k}{k!(n+k)!}
.
\end{equation}
Here $I_n(x)$ is the modified Bessel function and ${}_0F_1(n;x)$ is the
confluent hypergeometric functions. The functions $G_n(x)$ are positive for
positive $x$ and grow as $e^{2\sqrt{x}}$ for large $x$.  In addition,
$G^\prime_n(x)=G_{n+1}(x)$.  We rely on the identity
\begin{equation}
\int_0^\infty dx \,G_1(x) e^{-rx} = e^{1/r}-1
\qquad (r>0)
.
\end{equation}

The correlation function
$C(r,T) = e^{\alpha/(rT)}$ follows from
\begin{equation}
\tau_1(w,T)
=
\delta(w)
+
\frac{\alpha}{T} G_1\!\left(\frac{\alpha}{T} w \right) \theta(w)
.
\label{eq:24}
\end{equation}
Here $\theta(x)$ is the Heaviside step function. In turn,
\begin{equation}
\begin{split}
\tau(\mu,T) &= 4\pi \frac{d \tau_{1,c}(\mu,T)}{d\mu} 
\\
&= 4\pi \frac{\alpha^2}{T^2}
G_2\!\left(\frac{\alpha}{T} \mu \right)\theta(\mu)
+
4\pi \frac{\alpha}{T} \delta(\mu)
.
\end{split}
\label{eq:25}
\end{equation}
In \Eq{24}, $\delta(w)$ gives rise to the Polyakov loop, while in \Eq{25},
$\delta(\mu)$ accounts for the additional $1/r$ falloff in the connected part
of $C(r,T)$ for large separations.

Another case where the spectral function takes a closed form is that of
a simple model of the
type string tension plus Coulomb. Specifically, for the internal energy shift
\begin{equation}
\DU(r,T) = \sigma_R r -\frac{\alpha_R}{r} + U_0
,
\end{equation}
where $U_0$ is a constant. This form of $\DU(r,T)$ is suited to model the low
temperature regime of a Yang-Mills theory.  The thermodynamic relations
require then the entropy shift $\DS(r,T)$ to be a function of $r$ only. The
choice
\begin{equation}
\DS(r,T) = -\log(4\pi r \mu_0)
\end{equation}
allows to fulfill the spectral representation, $\mu_0$ being an arbitrary
scale.  The corresponding correlation function takes the form
\begin{equation}
C(r,T) = \frac{1}{4\pi r \mu_0} e^{-(\sigma_R r - \alpha_R/r + U_0 )/T}
.
\end{equation}
In this model $L(T)=0$. Such correlation then follows from
\begin{equation}\begin{split}
\tau(\mu,T) &= 
\frac{e^{ - U_0/T}}{\mu_0} \left[
\delta\Big(\mu-\frac{\sigma_R}{T}\right)
\\ & \ \ \ \ 
+
\frac{\alpha_R}{T} G_1\!\left(\frac{\alpha_R}{T}
\left(\mu - \frac{\sigma_R}{T}\right)
\right) 
\theta\left(\mu - \frac{\sigma_R}{T}\right)
\Big]
.
\end{split}
\label{eq:7}
\end{equation}

A Polyakov loop can be added in the correlation function without touching
$\tau(\mu,T)$ but in this case $\DU(r,T)$ and $\DS(r,T)$ are no longer given
by the previous expressions.

In \Eq{7}, the Dirac delta term represents the mass of the lowest-lying state,
a string of length $1/T$, which dominates the long distance tail of the
correlation in the string tension plus Coulomb model. On the other hand, the
Coulomb-induced term increases as an exponential of $\sqrt{\mu}$ and it should
saturate the large $\mu$ region of $\tau(\mu,T)$ of full QCD, albeit softened
by asymptotic freedom. This behaviour ensures the convergence of the measure
$\tau(\mu,T) e^{-r \mu}$ at large $\mu$ for $r>0$.

In Yang-Mills theory one should expect a gap in the $\mu$ mass spectrum below
the transition temperature, giving rise to a string tension. Indeed, a gap in
the spectrum above the vacuum state makes the integration in the spectral sum
rule to start at the mass $\mu_1>0$ of the lightest state and this term
dominates the large $r$ behaviour of the correlation function. Assuming that
$\mu_1$ is an isolated non vanishing point in the spectrum (the lightest state
is expected to be a single particle state) one finds, for large $r$,
\begin{equation}
C(r,T) \underset{r\to\infty}{\approx} C_0(T) \frac{e^{-\mu_1(T) r }}{4\pi r} 
\end{equation}
and the quantity $T\mu_1(T)$ can be interpreted as a (possibly $T$-dependent)
string tension. An increase in $T$ should translate into a continuous
quenching of the gap which becomes zero at the transition temperature. Above
this temperature $\tr_R\Omega^\dagger|0,T\rangle$ couples to the vacuum state
and a non vanishing Polyakov loop expectation value emerges.

If there were a gap also in full QCD one would have instead
\begin{equation}
C(r,T) \underset{r\to\infty}{\approx} C_0(T) \frac{e^{-\mu_1(T) r }}{4\pi r} +
L(T)^2 
\end{equation}
but still $T\mu_1(T)$ could be interpreted as a (once again $T$-dependent)
string tension.  The dominance of the Polyakov loop at large $r$ would display
the string breaking.

However, a gapless spectrum seems more likely for full QCD. A vanishing gap is
more suited to describe the deconfined phase due to the lack of a string
tension there. Then by continuity the gap must be zero for all temperatures,
since in QCD there is a crossover rather of a true phase transition
\cite{Aoki:2006we}. Also the fact that $L(T)$ is non zero for all temperatures
and representations suggests a vanishing gap. Nevertheless, even if strictly
speaking, the support of the spectral function $\tau(\mu,T)$ fills the
positive half-line, it is not excluded that, in the confined phase, this
function have a narrow peak around some $\mu_1(T)$. In this case one can
speculate that a precise definition of a QCD string-tension could still be
obtained as a pole in the $\mu$ complex plane.

\section{Renormalization group}
\label{sec:renormalization_group}

\subsection{Renormalization of the free energy}
\label{subsec:renormalization}

In the extraction of physical information from lattice calculations the
renormalization of the free-energy is a crucial issue. At distances shorter
than the thermal wavelength, $r T \ll 1 $, one expects that the medium plays a
minor role and, due to asymptotic freedom, perturbative QCD applies
~\cite{Brambilla:2010xn,Burnier:2009bk}. This requirement has been emphasized
in a series of insightful
works~\cite{Kaczmarek:2002mc,Zantow:2003uh}. A limitation is
that the necessary small values of $r$ might not be attainable in current
lattice settings. More recently, a full calculation with realistic dynamical
quark masses has been carried out for fundamental
sources~\cite{Borsanyi:2015yka,Bazavov:2016qod}. There, the lattice action is
renormalized at zero temperature, i.e., by following lines of constant
physics. The only new ultraviolet ambiguity introduced by the Polyakov loop
operators is a single $r$- and $T$- independent (although $R$-dependent)
additive constant in $F(r,T)$, i.e. $F(r,T) \to F(r,T) + c$, which in
principle can be fixed by setting $\DF(r_0,T_0)=F_0$ for conventionally chosen
$r_0$, $T_0$, and $F_0$.

We will review below the renormalization group equation (RGE) for the vacuum case
\cite{Ellis:1998kj,Shushpanov:1998ce} and the heavy quark situation
\cite{Chernodub:2010sq}. Our reanalysis is based on the entropy and introduces some minor but key modifications. The upshot is that while one can determine the entropy shift
from the RGE in the vacuum and single heavy quark case, one cannot do the same
in the case of heavy quark-antiquark sources separated a given distance.

\subsection{Renormalization group for the entropy}
\label{subsec:ren_group}

\subsubsection{Medium}

While the entropy or the free energy are of high theoretical interest,
experimentally one can only measure the specific heat. Thus we naturally
expect that this quantity be expressed in a manifestly renormalization group
invariant way. To fully appreciate this point let us consider the simplest
case, of a pure gauge theory which is specified by a renormalization scale
$\mu$ and a dimensionless coupling constant $g (\mu)$, at temperature T and
volume $V$. The partition function $Z$ and its relation to the free energy $F$
is defined as
\begin{eqnarray}
Z= e^{-F/T}= \int DA e^{-\int d^4 x \cL }
\end{eqnarray}
where the Lagrangian is given by 
\begin{eqnarray}
\cL = \frac14 (G_{\mu \nu}^a)^2  \,.
\end{eqnarray} 
From the standard thermodynamic relation we have the entropy
\begin{eqnarray}
S= - \frac{\partial F}{\partial T}= \frac{\partial }{\partial T} (T \log Z)
.
\end{eqnarray} 
As any dimensionless quantity, the entropy $S(T,V)$ must fulfill a functional
dependence involving just dimensionless quantities. If we take $g$, $\mu$, $T$
and $V$ as the relevant variables we have
\begin{eqnarray}
S(g,\mu,T,V) = \phi (g, \log (\mu /T), \log (\mu V^\frac{1}{3})) .
\end{eqnarray} 
The renormalization group invariance means that 
\begin{eqnarray}
\mu \frac{d}{d\mu} S(g,\mu,T,V) =0 
\end{eqnarray}
whence we obtain 
\begin{eqnarray}
T \partial_T S - 3 V \partial_V S = \beta(g) \partial_g S 
\,,
\end{eqnarray}
where the beta function is defined as 
\begin{eqnarray}
\beta(g) = \mu \frac{\partial g}{\partial \mu}
.
\end{eqnarray}
To evaluate the derivative with respect to $g$ we rescale the gluon field
$\bar A = g A$ so that the measure and the action scale as 
\begin{eqnarray}
DA = D\bar A /g^{N} \,, \qquad \cL = \frac{1}{4g^2} (\bar G_{\mu \nu}^a)^2 \,,
\end{eqnarray}
with $N$ the number of lattice points $N = N_t \times N_s^3 =  V/T
/a^4$ with $a$ the lattice spacing, and $\bar G_{\mu \nu} =
\partial_\mu \bar A_\nu- \partial_\nu\bar A_\mu + i [ \bar A_\mu, \bar
  A_\nu] $ is independent of $g$. Thus,
\begin{equation}\begin{split}
\partial_g S &= \partial_T ( T \partial_g \log Z) 
= \partial_T \left( T \frac{V}{T a^4} 
+ \frac{2}{g^3} V \langle (\bar G_{\mu\nu}^a)^2 \rangle_T   \right) \nonumber 
\\ &= 
\frac{V}{2g} \partial_T 
\langle (G_{\mu\nu}^a)^2 \rangle_T  
.
\end{split}\end{equation}
Defining the trace of the energy momentum tensor as
\begin{eqnarray}
\Theta = \frac{\beta(g)}{2g} (G_{\mu\nu}^a)^2
\end{eqnarray}
we get finally
\begin{eqnarray}
\partial_T (E-3 P V)
= \partial_T \left[T \int  d^4 x \langle \Theta \rangle _T \right]
.
\end{eqnarray}
In the case of infinite volume we just get the energy density
$\epsilon= E/V$ and $T \int d^4 x= \int d^3 x = V $ and the volume
factors out
\begin{eqnarray}
\partial_T (\epsilon-3 P )
= \partial_T \langle \Theta \rangle_T 
.
\end{eqnarray}
Therefore, integrating from $0$ to $T$ one gets 
\begin{eqnarray}
\epsilon-3 P 
= \langle \Theta \rangle_T - \langle \Theta \rangle_0 
,
\end{eqnarray}
where we have assumed that $E,P \to 0$ for $T \to 0$. This follows from the
low temperature partition function behaviour $O( e^{-m_G/T})$ where $m_G$ the
mass of the lightest glueball. The derivation of
Ref.~\cite{Ellis:1998kj,Shushpanov:1998ce} starts already with the partition
function in terms of the scaled fields $\bar A$ and thus the final equation
does not include the subtracted contribution at zero temperature, which has to
be added by hand (see also \cite{Chernodub:2010sq}).

In the full QCD case we have to add quark and anti-quark fields  
\begin{eqnarray}
Z= \int DA D\bar q D q  \,  e^{-\int d^4 x \cL (x)} = e^{-F/T}
,
\end{eqnarray} 
where the QCD Lagrangian for $N_f=3$ flavours
$u,d,s$ reads, in terms of the re-scaled gluon field $A_\mu = \sum_a
A_\mu^a T_a $ with $\tr (T_a T_b) = \delta_{a,b}/2 $ and $\bar A_\mu^a
= g A_\mu^a $,
\begin{eqnarray} 
\cL (x) = \frac{1}{4 g^2} (\bar G_{\mu \nu}^a)^2 + \sum_{q=u,d,s} \bar q
(i \slash \!\!\!\!\! D + m_q ) q
.
\end{eqnarray} 
Renormalization group invariance requires the inclusion of the mass terms by
assuming an extra dependence on the dimensionless variable $\log (\mu
/m_q(\mu))$, yielding
\begin{eqnarray}
\mu \frac{dS}{d\mu}  &=&  \beta(g) \frac{\partial S}{\partial g} 
- \sum_q  m_q (1+\gamma_q)  \frac{\partial S}{\partial m_q} 
- T \frac{\partial S}{\partial T}
+ 3 V \frac{\partial S}{\partial V}  \nonumber \\ 
 &=& 0
, 
\end{eqnarray}
with the beta function and the mass anomalous dimension given by
\begin{eqnarray}
 \beta(g) = \mu \frac{dg}{d\mu}\, ,  \qquad \gamma_q(g)= -\frac{d \log m_q}{d
   \log \mu} 
\, .
\end{eqnarray}
We obtain the same formulas as above with the energy momentum tensor
defined as
\begin{eqnarray}
\Theta \equiv \Theta^\mu_{\mu} = \frac{\beta (g)}{2g} (G^a_{\mu \nu})^2 
+ \sum_q m_q (1+\gamma_q) \bar q q 
.
\end{eqnarray} 
Here, in the low temperature limit $E$ and $P$ are saturated by the free pion
gas which again provides a vanishing contribution of the type
$\Or(e^{-m_\pi/T})$.

\subsubsection{Heavy Source}

For an operator $\cO$ the thermal expectation value is defined as
\begin{eqnarray}
\langle \cO \rangle_T = 
\frac{\int DA  \, \cO \, 
e^{-\int d^4 x \cL (x)}}{\int DA \, e^{-\int d^4 x \cL (x)}}
.
\end{eqnarray} 
A qualification is in order here. Generally, not every thermal expectation
value corresponds to a ratio of partition functions. Strictly speaking a
partition function requires a spectral decomposition of the form $Z= \sum_n
g_{n} e^{-E_n/T}$ where $g_n$ is a non-negative integer. We will only consider
operators which indeed correspond to partition functions (in particular $\cO$
must be a dimensionless projector operator)
\begin{eqnarray}
\frac{Z_\cO}{Z}= e^{-\Delta F_\cO/T} 
\end{eqnarray}
where we have introduced the free energy shift, 
\begin{eqnarray}
\Delta F_\cO = F_\cO - F
.
\end{eqnarray}
To this free energy shift there corresponds an entropy shift,
\begin{eqnarray}
\Delta S_\cO = - \frac{\partial \Delta F_\cO}{\partial T}
.
\end{eqnarray}
The renormalization group equation becomes 
\begin{eqnarray} 
T \partial_T \Delta S_\cO - 3 V  \partial_V \Delta S_\cO
= \partial_T \left[T \int_V d^4 x \langle \Theta \rangle_{\cO,T} \right]
\end{eqnarray} 
where 
\begin{eqnarray}
\langle \Theta \rangle_{\cO,T} =  \frac{\langle  \Theta \cO \rangle_{T}}
{\langle \cO \rangle_T} - \langle \Theta \rangle_{T}
\,.
\end{eqnarray}

We can take the operator $\cO$ as the Polyakov loop in any representation
$R$ and, in the infinite volume case, the volume dependence term drops
out. Furthermore, for this choice of $\cO$ the thermal expectation value
corresponds to the ratio of two true partition functions~(\Eq{1}).  We will
work in the static gauge, in which the Polyakov loop reads $\tr_c (\Omega
(\vec x)) = \tr_c (e^{i g A_0(\vec x)/T}) $ ($ \tr_c {\bf 1}=N_c$) and we take
conventionally $\vec{x}=0$.  In this case the RGE reads
\begin{eqnarray} 
T \partial_T \Delta S_R = \partial_T \left[T \int d^4 x \langle \Theta
  \rangle_{R,T} \right]
.
\end{eqnarray} 
This formula provides the specific heat for placing a colour charge in
the representation $R$ into the hot medium. Remarkably this equation
allows to uniquely determine the entropy after specifying its value at
a given temperature. 

In the limit of small temperatures we have
\begin{eqnarray}
\Delta S_R (0) = \log D_R 
,
\end{eqnarray}
where $D_R$ is the ground state degeneracy in the subspace with a colour
source in the representation $R$ at $\vec{x}=0$. In particular, if the source
is in the fundamental representation the lowest state is obtained by screening
the source by a single light anti-quark, hence $D_R=2N_f$ for $N_f$
mass-degenerated light quark flavors (the factor $2$ coming from the two spin
states of the anti-quark) ~\cite{Megias:2012kb}. The $N_c$ different
colour states of the source are combined with those of the light quark to
form a colour singlet, so this degree of freedom does not add to the entropy.

In the opposite limit of large temperatures a partition function is just the
dimension of the Hilbert space of the system. Hence $Z_R\sim d_R Z_0 $ where
$d_R$ is the dimension of the representation $R$, thus
\begin{eqnarray}
\Delta S_R (\infty) = \log d_R
. 
\end{eqnarray}
This expresses the fact that at very high temperatures the colour state of the
source is not effectively correlated with the medium, so it just counts
additively for the entropy.\footnote{Recall that a colour source has no other
  degrees of freedom than colour, by definition. A heavy quark serves as a
  source, not only because it does not exchange kinetic energy but also
  because its spin state is fully decoupled and can be disregarded.} [For the
  free energy defined from the Polyakov loop normalized as
  $\langle\tr\Omega\rangle/d_R$, this result takes the form $\Delta S_R
  (\infty) = 0$.]

\subsubsection{Heavy source correlator}

In the case of a Polyakov loop correlator there appears the separation between
conjugate sources, $r$, which enters in the RGE by including the dimensionless
quantity $\log (\mu r)$ and yielding the replacement $T \partial_T \to T
\partial_T -r \partial_r$, thus
\begin{eqnarray}
T \partial_T \Delta S_{\bar R \otimes R} - r \partial_r \Delta S_{\bar R \otimes R}
 = \partial_T \left[T \int d^4 x \langle \Theta \rangle_{\bar R \otimes R,T} \right]
.
\label{eq:54a}
\end{eqnarray} 
It is noteworthy that the previous equation is free from UV divergences,
although the integrand itself can display UV divergences at the heavy source
points. This is at variance with a RGE for the free energy
\cite{Chernodub:2010sq}.

\Eq{54a} is a standard first order partial linear differential equation which
can be solved by the method of characteristics.  Assuming that the r.h.s. is
known, the equation provides the variation of $\Delta S_{\bar R \otimes R}
(r,T) $ along paths $rT={\rm constant}$ in the $(r,T)$ plane. Specifically
\begin{eqnarray}
T \partial_T S - r \partial_r S = \phi(r,T)
\end{eqnarray}
is equivalent to
\begin{eqnarray}
S(r,T) = S(rT/T_0,T_0) + \int_{T_0}^T  \phi(r T/T', T') \frac{dT'}{T'} 
.
\label{eq:56}
\end{eqnarray}
The determination of $S(r,T)$ from the RGE can then be achieved from the
knowledge of the entropy along a line visiting all the $rT$ values.

The explicit solution \Eq{56} applies immediately if the entropy is known for
all $r$ at some reference temperature $T_0$. For instance in the $ \bQ Q$
case, at low temperatures we expect
\begin{equation}
 \lim_{T \to 0} \Delta S_{\bQ Q}(r, T) = \left\{\begin{matrix}
 0 & r < r_c \\ 2 \log (2 N_f) & r > r_c
\end{matrix}\right.
\end{equation}
where $r$ is kept constant as $T\to 0$, and $r_c$ is the string breaking
distance.  This region of small $T_0$ is beyond a perturbative calculation. On
the other hand, at high temperatures we expect a free theory, with
\begin{eqnarray}
\lim_{T \to \infty} \Delta S_{\bQ Q}(r, T) = 2 \log (N_c)  
.
\end{eqnarray}
In this $T_0\to 0$ limit, $S(rT/T_0,T_0) $ lies in the perturbative region
since $r_0=rT/T_0$ is also small. Nevertheless, likely, any perturbative
corrections will be exponentially inflated by the RGE, rendering the
determination of $S(r,T)$ for finite $(r,T)$ unreliable.

\subsection{Perturbation theory and RGE improvement}


\subsubsection{Heavy source in the medium}

One useful application of the RGE is the derivation of constraints on
perturbative results. For instance, the expectation value of the Polyakov loop
has been computed to $O(g^4)$ in pQCD in
\cite{Gava:1981qd,Brambilla:2010xn,Burnier:2009bk} and to $O(g^5)$ in
\cite{Berwein:2015ayt}. To this order, the structure of this quantity is
\begin{equation}\begin{split}
\frac{1}{N_c}L(T) &= 1 + c_0 g^3 + (c_1 + d_1 \log g ) g^4 
\\ &
\quad
+ \left(c_2 + e_2 \log\left(\mu/2\pi T \right)\right) g^5 
+ \Or(g^6)
,
\end{split}\end{equation} 
hence, for the entropy at NNLO
\begin{equation}\begin{split}
\Delta S_Q &= 
c_0  g^3
+ \left( c_1 + d_1 \log g \right) g^4
\\ & \quad
+ \left(
c_2 - e_2 + e_2 \log \left(\mu /2\pi T\right) 
\right) g^5
+ \Or(g^6)  
.
\end{split}\end{equation}

The RGE then requires $ e_2 = 3 \beta_0 c_0$, where
\begin{equation}
\beta(g) = - g^3 \sum_{n\ge 0} \beta_n g^n
.
\end{equation}

A determination of the specific heat 
\begin{equation}
\Delta c_Q \equiv T\partial_T \Delta S_Q
\end{equation}
by direct derivation with respect to $T$ with fixed $\mu$ would produce just
the LO result (of $\Or(g^5)$). However, such a calculation would disregard the
fact that some of the $T$-dependence at higher orders is fixed by the known
lower orders through the RGE. The RGE guarantees that to any given order we
can differentiate $S_Q$ (with fixed $\mu$) and then take $\mu=\mu(T)$
(e.g. $2\pi T $) or the other way around. The latter method recovers the NNLO
result for $c_Q$. Alternatively, one can work with fixed $\mu$ and use the RGE
relation
\begin{equation}
T\partial_T \cO  = \beta(g) \partial_g \cO
,
\end{equation}
which holds for any observable in which $T$ is the only physical scale.
This gives
\begin{equation}\begin{split}
c_Q &= 
-3 \beta_0 c_0  g^5
-
\left(\beta_0 (4 c_1 + d_1 + 4 d_1 \log g ) + 3 \beta_1 c_0 \right) g^6
\\
& \quad
+ \Big(
5 \beta_0 \left(- c_2 + e_2 - e_2 \log\left(\mu/2\pi T\right) \right)
\\ & \quad 
- \beta_1 (4 c_1 + d_1 + 4 d_1 \log g )
-3 c_0 \beta_2
\Big) g^7
+\Or(g^8)
.
\end{split}\end{equation}
Note that the $\beta$ function to NNLO is needed.

\subsubsection{Heavy source correlator}

The perturbative calculation of the Polyakov loop correlator was done by
Nadkarni~\cite{Nadkarni:1986cz} in the regime $T\gg 1/r \sim m_D$ ($m_D$ is
the Debye mass) yielding a function of $ r T$ for the connected piece. Such
dependence vanishes under the RGE for the entropy, up to higher orders in
perturbation theory stemming from the energy-momentum tensor
contributions. This illustrates explicitly that the RGE of the entropy shift
is not sufficient for the full reconstruction of the correlator.  In the
calculation of Ref. \cite{Brambilla:2010xn}, in the regime $1/r \gg T \gg
m_D$, a non vanishing contribution to the RGE is explicitly obtained, involving
again higher orders. Therefore, in order to test these perturbative results, a
different object should be computed, namely, the energy momentum density of
two charge conjugated heavy sources at finite temperature. To our knowledge
such a calculation has never been carried out in any form in the literature.

\section{Heavy $\bQ Q$ free energy shift in the confined phase}
\label{sec:confined_phase}

On physical grounds one expects that below the phase transition the
free-energy shifts should be expressed in terms of hadronic colour singlet
states, but the precise mathematical formulation of this expectation has never
been made clear. We will address in this section the hadronic representation
of the heavy $\bQ Q$ free-energy shifts, based on our findings of
Sec.~\ref{sec:SpectralRepresentation}. A hadronic representation for the
Polyakov loop was already derived in~\cite{Megias:2012kb} based on first
principle arguments of QCD. The results of this section generalize that
study.

\subsection{$\bQ Q$ spectrum and string breaking}
\label{subsec:spectrum}

At zero temperature and for fundamental sources, the concept of string tension
has played a major role in the formulation and understanding of confinement
and colour gauge invariance in the pure Yang-Mills theory. In the zero
temperature limit one has for large distances $ \DF (r,0) \sim \sigma r $. In
QCD however, the string between fundamental charges breaks generating a $\bq
q$ pair from the vacuum which subsequently decays into hadronic states, and so
instead $ \DF (r,0) \sim 2 \Delta $ where $\Delta$ is the mass of the
lowest-lying bound state (a hybrid heavy-light meson). The breaking of the
string has been studied on the lattice~\cite{Bali:2005fu} where the
avoided crossing between the $\bQ Q$ and the $\bM M \equiv (\bQ
q) (\bq Q) $ channels was observed, which is familiar from molecular physics in the Born-Oppenheimer
approximation~\cite{landau1965quantum}. More generally, the $\bQ Q$ state can
decay into any of the many excited states of the meson spectrum, or the baryon
spectrum, as long as they have the same quantum numbers as the $\bQ Q$
system.\footnote{The mechanism of decaying into baryons implies the creation
  of two pairs of light quarks $\bq q$, leading to the formation of two
  heavy-light baryons with one heavy quark.}

To analyze the $\bQ Q$ free-energy lattice results in the confined phase we
will apply the approximate $T$-spectral representation in \Eq{4} and the exact
$r$-spectral representation in \Eq{L2}. Further, we will consider the
coupled-channel Hilbert space spanned by $\bQ Q$ (a single state representing
the colour sources joined by a string in its ground state) and by pairs $\bH
H$ of heavy hadrons.  Here $H$ is a colour source screened with light quarks
and gluons to form a singly-heavy hadron (meson or baryon), either low-lying
or excited. As in \cite{Megias:2012kb}, half of the heavy hadron spin states
are spurious, since the colour source has no spin. When this source is
simulated with a very heavy quark, its spin decouples due to QCD heavy quark
spin symmetry \cite{Isgur:1989vq,Neubert:1993mb}.  The energies are modeled as
\begin{equation}\begin{split}
V_{\bQ Q} (r) &= -\frac{4}{3}\frac{\alpha}{r} + \sigma r  + c \,, 
\\
V^{(n,m)}_{\bH H}(r) &= \Delta^{(n)}_{\bH}+ \Delta^{(m)}_{H} \,, 
\end{split}
\label{eq:string-breaking-bis}
\end{equation}
where $ \Delta^{(n)}_{H} = M^{(n)}_H - m_{Q}$ ($n$-th heavy-light hadron mass
minus heavy quark mass).  An additive constant $c$ has been included in
$V_{Q\bar{Q}}(r)$ to account for the ambiguity in the renormalization of the
lattice data (see Sec.~\ref{subsec:renormalization}) In addition, a $\bQ Q
\leftrightarrow \bH H$ transition potential $W$ can be present.

In principle one should take the infinite quark mass limit ($m_Q\to \infty$)
and compute the corresponding spectrum. This is an ambitious program and as a
guide we will content ourselves with using existing extensive quark model
calculations containing the main essential features describing singly heavy
hadrons (mesons and baryons) and relativity, which proves crucial to account
for excited states. Obviously this represents an approximation but these
models already produce a trace anomaly for $u$, $d$, $s$ quarks, below the
phase transition, which can hardly be distinguished by the conventional hadron
resonance gas using the listed PDG values \cite{Arriola:2014bfa}. Besides,
one can assess the corresponding uncertainty by comparing the $c$-hadron versus
the $b$-hadron spectra and this is in accordance with previous analyses for
the Polyakov loop, cf.~\cite{Megias:2012kb,Megias:2012hk}.
Moreover, we will disregard gluonic string-like excitations as we expect them
to have a gap of the order of $\sqrt{\sigma} \sim 400\MeV$.

\begin{figure}[t]
\begin{center}
\epsfig{figure=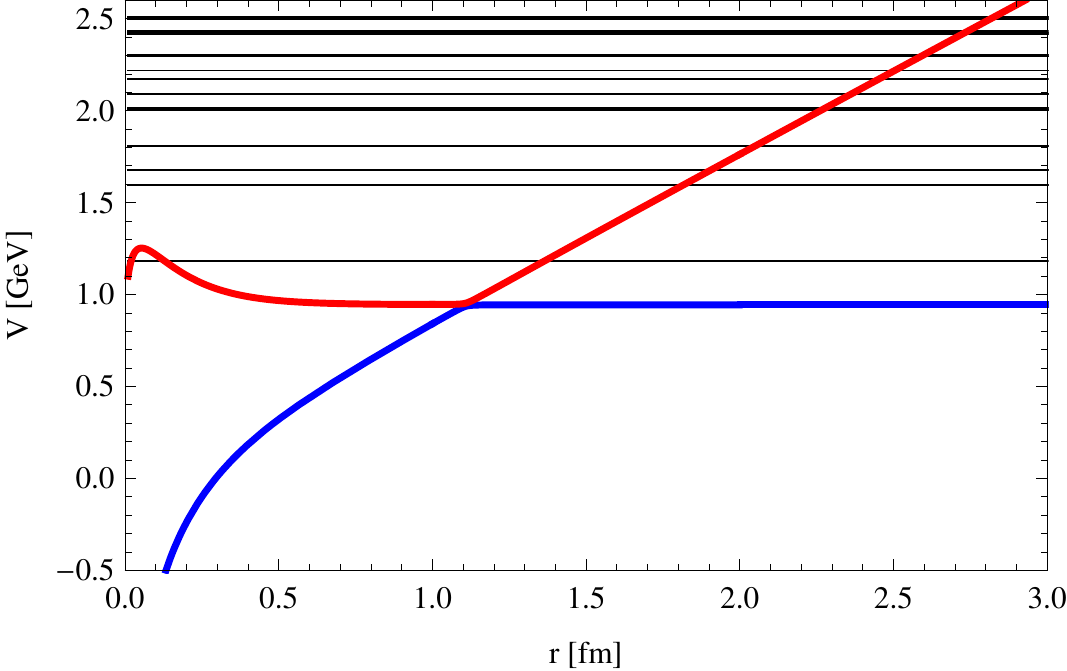,height=50mm,width=85mm}
\epsfig{figure=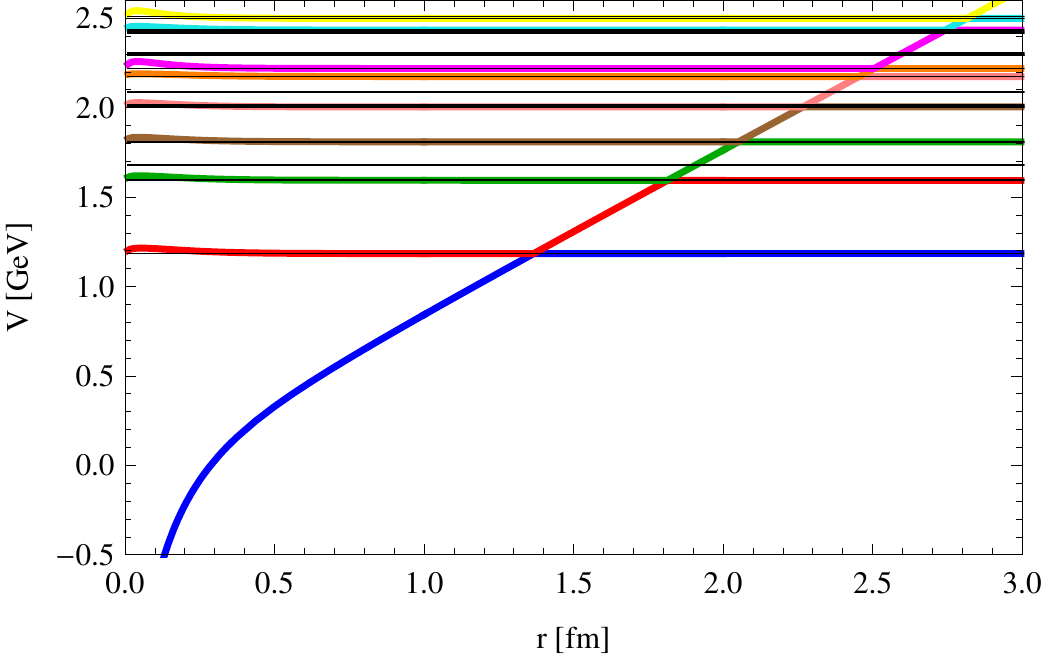,height=50mm,width=85mm}
\end{center}
\caption{The spectrum as a function of distance and the (avoided) level crossing
  structure for the considered string-meson mixing scenarios. Single mixing
  (upper panel) and multiple mixing (lower panel) for the model which includes
  the Isgur states for heavy-light mesons with a charm quark. We consider the
  avoided crossing (thick lines) shown in Fig.~\ref{fig:plotW} (up).}
\label{fig:plotcrossings}
\end{figure}
In Fig.~\ref{fig:plotcrossings} we illustrate the situation for $\sigma=(0.425
\GeV)^2$ and $\alpha = \pi/16$, and using the spectrum of heavy-light hadrons
with a charm quark from the Godfrey-Isgur relativized quark model
(RQM)~\cite{Godfrey:1985xj} up to $\Delta_B = 3.19 \GeV$. As mentioned, a
source of error is that the heavy quarks in nature have a finite mass. In
order to estimate this, we have used as well the spectrum of heavy-light
hadrons with a $b$ quark from the very same model. Besides reproducing
accurately de PDG level density, the RQM satisfactorily describes the Polyakov
loop below the crossover and, for the heavy-light system with charm or bottom
quarks, yields an exponentially growing spectrum $\rho \sim e^{\Delta_B/T_H}$
with Polyakov-Hagedorn temperature $T_{H,c}=210 \MeV$ \cite{Arriola:2013jxa}.
Before mixing, all $\bH H$ levels cross with the quenched $\bQ Q$ potential
(note the relatively large gaps between lower $\bH H$ levels mixing with the
$\bQ Q$ channel). The avoided crossings correspond to $\bH H$ states with the
same $J^{PC} $ quantum numbers as the $\bQ Q$, which are also highlighted in
Fig.~\ref{fig:plotcrossings} for the RQM using a mixing $W$ to be discussed
subsequently.

\subsection{Hadronic representation for the free energy}
\label{subsec:hadronic_representation}

Relying on the $T$-spectral representation and adding up all coupled-channel
states, {\em in the absence of mixing} one finds~\cite{Arriola:2014bfa}
\begin{equation}
e^{-\DF(r,T)/T}  = e^{-V_{\bQ Q}(r)/T} + \left(\sum_n e^{-\Delta^{(n)}_{H}  /T}\right)^2 
.
\label{eq:Fave2}
\end{equation}
The last term is just $L(T)^2$~\cite{Megias:2012kb}. The simple form of this
term follows from neglecting any interaction between the two heavy hadron
states (\Eq{string-breaking-bis}).

\begin{figure}[t]
\begin{center}
\epsfig{figure=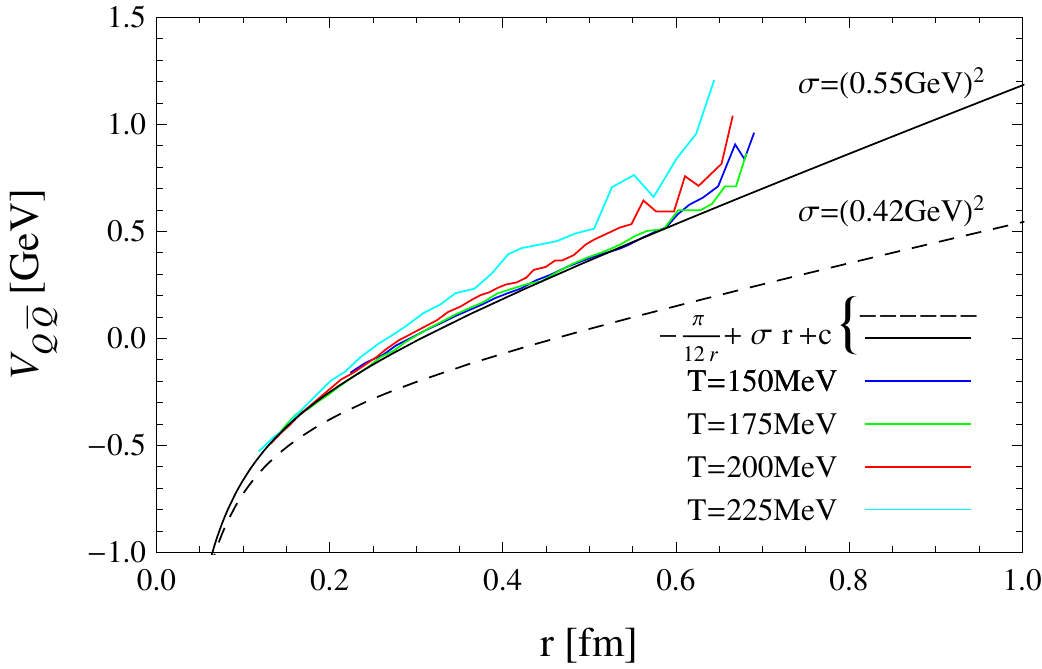,height=60mm,width=85mm}
\epsfig{figure=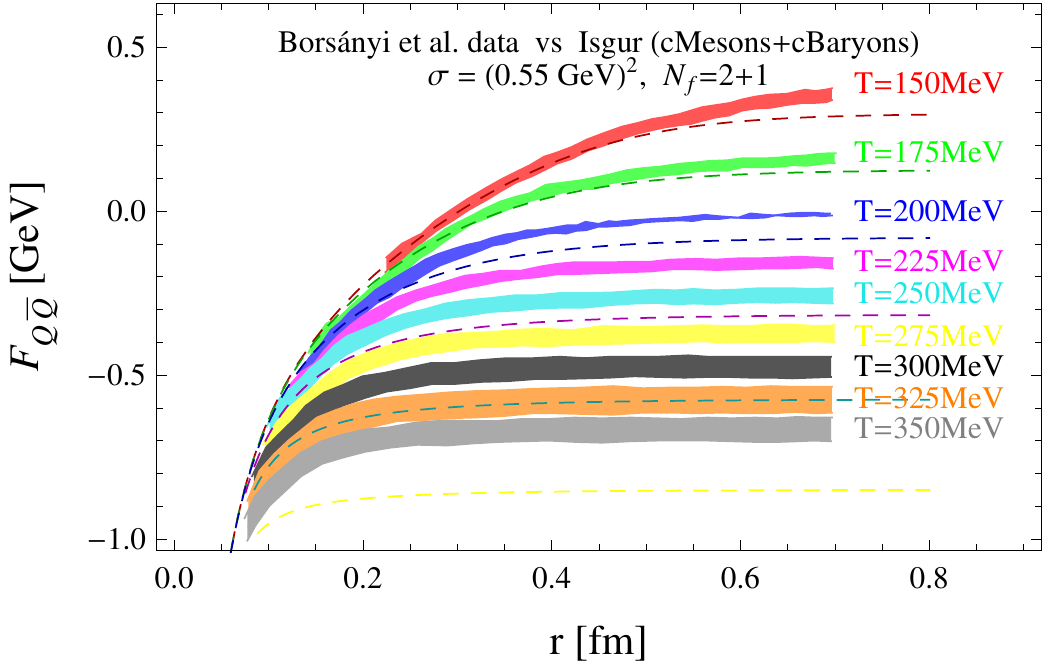,height=50mm,width=85mm}
\end{center}
\caption{Upper panel: Heavy $\bQ Q$ potential at zero temperature as a
  function of the separation, obtained by solving \Eq{Fave2} and taking the
  lattice data shown in the lower panel. The continuous (resp. dashed) black line is
  the result of the Cornell potential, \Eq{string-breaking-bis}, with
  $\sigma=(0.55 \GeV)^2$ (resp. $\sigma=(0.42 \GeV)^2$) and $c=-0.3
  \GeV$.  Lower panel: Heavy $\bQ Q$ free energy as a function of the
  separation. The bands are the lattice data for $N_f=2+1$ taken from
  Ref.~\cite{Borsanyi:2015yka}. The dashed lines represent the result by using
  \Eq{Fave2} with the spectrum of heavy-light mesons and baryons with a charm
  quark from the Isgur model of Ref.~\cite{Godfrey:1985xj}.  }
\label{fig:Vqq}
\end{figure}
Using the data of Ref.~\cite{Borsanyi:2015yka} to extract $\DF(r,T)$ and
$L(T)$, one can determine $V_{\bQ Q}(r)$ by inverting \Eq{Fave2}.  This yields
a $T$-independent potential for $r T \ll 1 $ but with a string tension $\sigma
= (0.55 \GeV)^2$, almost twice the conventionally accepted value $(0.42
\GeV)^2$. This is shown in Fig.~\ref{fig:Vqq} (up). This disturbing result
suggests that determining the string tension or renormalizing the
free-energy-shift from short distances (at least in lattice QCD with realistic
quark masses and current lattice spacings) may contradict the long distance
and well established charmonium phenomenology, based on quenched
determinations of the string tension.

In Fig.~\ref{fig:Vqq} (down) we display the lattice results
of~\cite{Borsanyi:2015yka} for the heavy $\bQ Q$ free energy along with the
calculation from \Eq{Fave2} using the Isgur model with charm quarks and
$\sigma=(0.55 \GeV)^2$. A better agreement with lattice data is achieved
mainly for the lowest temperatures. The disagreement at higher temperatures is
in part motivated by the failure of the hadronic representation of the
Polyakov loop to describe the lattice data at higher temperatures,
cf.~\cite{Megias:2012kb}.

In order to estimate finite mass effects, the above determination of $F(r,T)$
from \Eq{Fave2} can be repeated using instead $b$ quarks in the Isgur
spectrum. One finds a maximum relative change in this quantity of $0.22\%$ for
$T=150 \MeV$ and of $6.17\%$ for $T=175 \MeV$. Such maximum deviations take
place at large separations as they reflect the slightly different Polyakov
loop predicted by the two versions, already noted in
Ref.~\cite{Megias:2012kb}. The variation is larger at the highest temperature,
where the model is less reliable. As we explain subsequently, in the
extraction of precise quantities like the mixing potential in
Sec.~\ref{subsec:avoided_crossings}, we take the Polyakov loop directly from
lattice data rather than from a hadronic model. This fixes the large
separation regime, hence finite mass effects are restricted to intermediate
distances, and they are significantly smaller than the above estimates.

Remarkably, \Eq{Fave2} is fully consistent with the exact $r$-representation,
\Eq{L2}. However, when channel mixing is allowed and the modified eigenvalues
are introduced in the $T$-spectral representation, the admissible mixings get
constrained by the concavity relations in \Eq{8}. For instance, in the low
temperature limit the lowest eigenvalue dominates and this implies the
conditions $E_0^\prime(r) \ge 0 $ and $E_0^{\prime\prime}(r) \le 0 $ (no such
direct constraint applies to the excited states). The two-level setting in
\Eq{V2states} (below) illustrates that $W$ cannot be arbitrary: for large
enough $W>0$ (the sign is conventional) $E_0(r)\sim -W(r)$ and so the
concavity relations require $W^\prime \le 0$ and $W^{\prime\prime} \ge 0$.

\subsection{Avoided crossings}
\label{subsec:avoided_crossings}

When mixing is switched on, we still use an additive model for
$V_{\bH H}^{(n,m)}(r)$, i.e., interactions between the two heavy hadrons
are neglected, but non vanishing matrix elements arise between different
$\bH H$ pairs, as well as transitions between the $\bH H$ states and the
$\bQ Q$ state. The simple form in \Eq{Fave2} no longer holds since the
Hamiltonian is no longer diagonal in the coupled-channel space.  If $V(r)$
denotes the Hamiltonian (a matrix in coupled-channel space) the formula with
mixing becomes
\begin{equation}
  e^{-F(r,T)/T} = \tr(e^{-V(r)/T}) = \sum_\alpha e^{-E_\alpha(r)/T}
,
\end{equation}
where $E_\alpha(r)$ denote the corresponding eigen-energies. 

To study the phenomenon of avoided crossings in closer detail and paralleling
the treatment in \cite{Bali:2005fu}, let us introduce a mixing just between
the ${\bQ Q}$ state and the lowest-lying $\bH H$ state
\begin{equation}
V(r) = 
\left(
\begin{array}{ccc}
-\frac{4\alpha}{3r} + \sigma r  & W(r) &  \\
W(r)  &  2\Delta & 
\\
 &  & \ddots
\end{array}
\right)
. 
\label{eq:V2states}
\end{equation}
We only display the two coupled levels of $V(r)$, this matrix being diagonal
for the remaining states.  To be more precise, the lowest lying $\bH H$ level
contains several spin-flavor states.  To match the quantum numbers of the $\bQ
Q$ state, we only couple the flavorless and spinless $\bH H$ combinations.
The uncoupled $\bH H$ states give a contribution as in \Eq{Fave2} and their
role is to saturate the Polyakov loop. Therefore their detailed eigen-energies
are not needed. The contribution obtained from diagonalization of the two
first states in $V(r)$ in \Eq{V2states} is computed explicitly and the
remainder is fixed so that the Polyakov loop value is reproduced.

\begin{figure}[t]
\begin{center}
\epsfig{figure=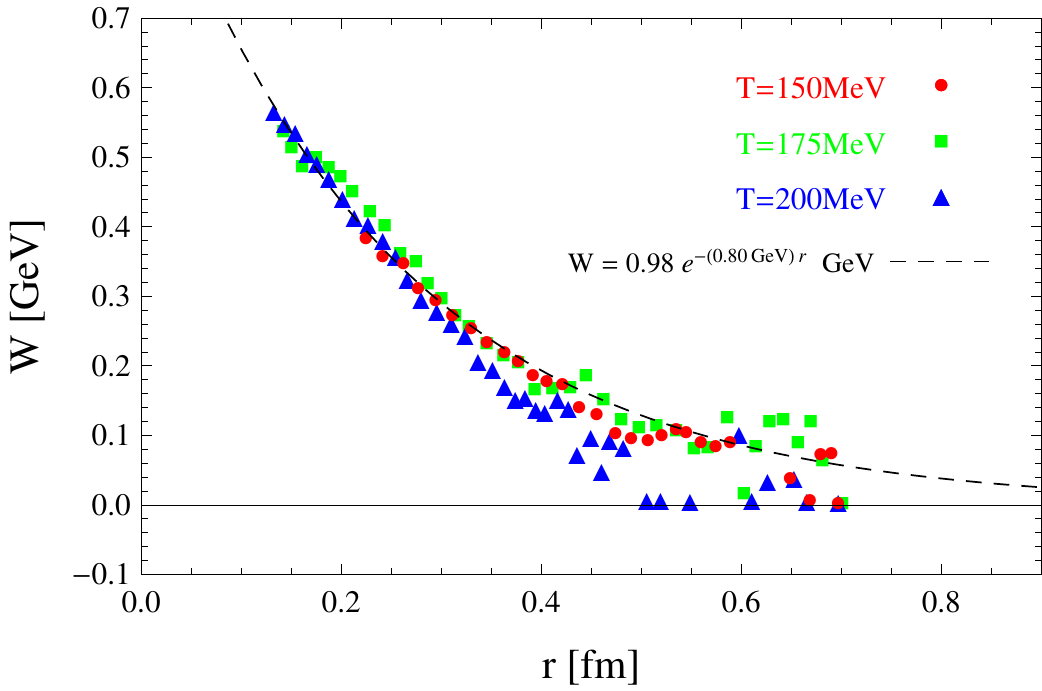,height=60mm,width=85mm}
\epsfig{figure=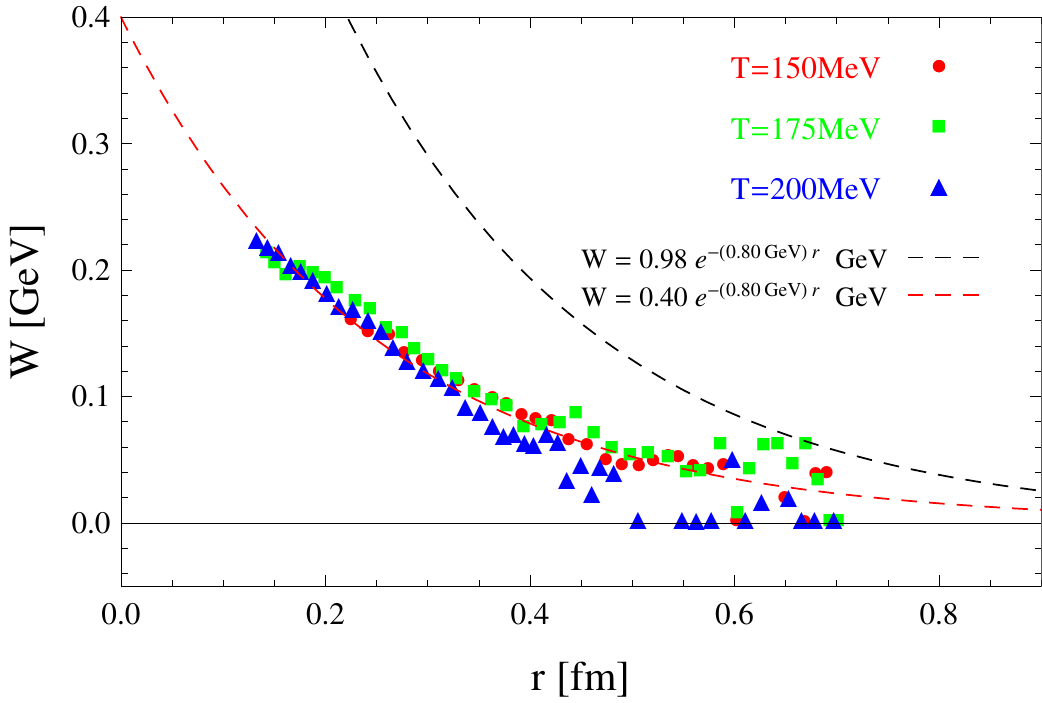,height=60mm,width=85mm}
\end{center}
\caption{Upper panel: Value of the mixing $W$ as a function of separation with
  the two states model of \Eq{V2states}. The dots result from a fit of the
  heavy $\bQ Q$ free energy lattice data of Ref.~\cite{Borsanyi:2015yka} with
  $\sigma = (0.42\GeV)^2$ and not assuming any functional form for $W(r)$. The
  dashed line corresponds to the best fit to the lowest temperature data
  assuming the functional form $W= g e^{-mr}$ with $\sigma$, $g$, and $m$ as
  parameters. The two methods agree and $W(r)$ is temperature
  independent. Lower panel: $W(r)$ as a function of separation with a model
  including mixing between string and all the heavy-light charmed meson states
  of the Isgur model. The effect of the mixing with the new states is to
  reduce the strength of the potential $W$, but not its exponential
  behaviour.}
\label{fig:plotW}
\end{figure}

This simplest mixing model allows to determine $W(r)$ {\em point by point}
from $\DF(r,T)$. A crucial consistency check of the approach is that $W$ must
be $T$-independent.
Clearly, our string-hadronic model will not be valid above the critical
temperature except for very small separations. Therefore, we expect that
extraction of $W(r)$ from the data will reflect this short distance medium
independence.  
As we can see from Fig.~\ref{fig:plotW} (up) the short
distance behaviour is pretty much independent of the temperature {\em
  provided} we take the standard value for the string tension, $\sigma =
(0.42\GeV)^2$. Moreover, in the common range for temperatures $T=150, 175, 200
\MeV$ an exponential behaviour is obtained (at least above $0.13\fm$), in
agreement with the functional form proposed in Refs.
~\cite{Drummond:1998ir,Drummond:1998he,Drummond:1998ar}. A fit with $W=g
e^{-mr}$ to the lattice data at $T=150\MeV$ with $\alpha=\pi/16$ and
$\Delta=472\MeV$ (the lowest Isgur state with a charm quark) produces
\begin{eqnarray}
&& \sigma = (0.424(14)\GeV)^2 \,, \qquad g=0.98(47)\GeV \,, \nonumber \\
&& m=0.80(38)\GeV \,, \label{eq:sgm}
\end{eqnarray}
with $\chi^2/\textrm{dof}=0.031$. Parameters $g$ and $m$ are highly
positively correlated, with a correlation $r(g,m) = 0.983$. 

For $\Delta= 553\MeV$, which corresponds to the lowest Isgur state with a
bottom quark, one obtains $\sigma = (0.424(14) \GeV)^2$, $g = 1.02(47) \GeV$,
and $m = 0.77(38) \GeV$, with $\chi^2/{\rm dof}= 0.031$ and $r(g,m)=0.983$.

We have checked that including more states in the mixing just reduces the
strength of the transition potential $W$ but does not change the exponential
behaviour, cf. Fig.~\ref{fig:plotW} (down).

Note that the highest temperature used in this analysis, $T = 200\MeV$ is
clearly above the physical value of $T_c$ in QCD which is around $150\MeV$.
The value of $T_c$ in Ref. \cite{Borsanyi:2015yka} is obtained in
\cite{Aoki:2006br,Aoki:2009sc} from chiral susceptibilities, namely,
$T_c=146(2)(3)\MeV$, in quantitative agreement with the HotQCD collaboration
\cite{Bazavov:2011nk}. On the other hand the approach presented in this
section is expected to be valid only in the confined regime.  In either case,
it is reassuring that such an approach can describe lattice data at
sufficiently short distances so close to the phase transition.

The form of the mixing potential is qualitatively similar to that found in
Ref.~\cite{Bali:2005fu} where the correlators of the matrix were computed at
zero temperature, but for pion masses around $700 \MeV$. Since the lattice
data used in our determination of $W(r)$ were obtained for physical hadron
masses we do not expect the observed exponential fall-off in $W(r)$ to
correspond to pion exchange. Indeed, the iterated mechanism, $\bH H \to \bQ Q
\to \bH H$ represents meson-exchange, and conversely, any meson-exchange must
involve some component of $\bQ Q$ as an intermediate state, but there is no
compelling reason to expect that the lighter mesons should be dominant in the
transition potential $W(r)$ in the range of distances considered. The value
$m=800\MeV$ is in the range of a typical meson mass with no special
mass-reducing mechanism (e.g., chiral symmetry in the pion case) at work.

\subsection{Entropy and specific heat}

\begin{figure}[t]
\begin{center}
\epsfig{figure=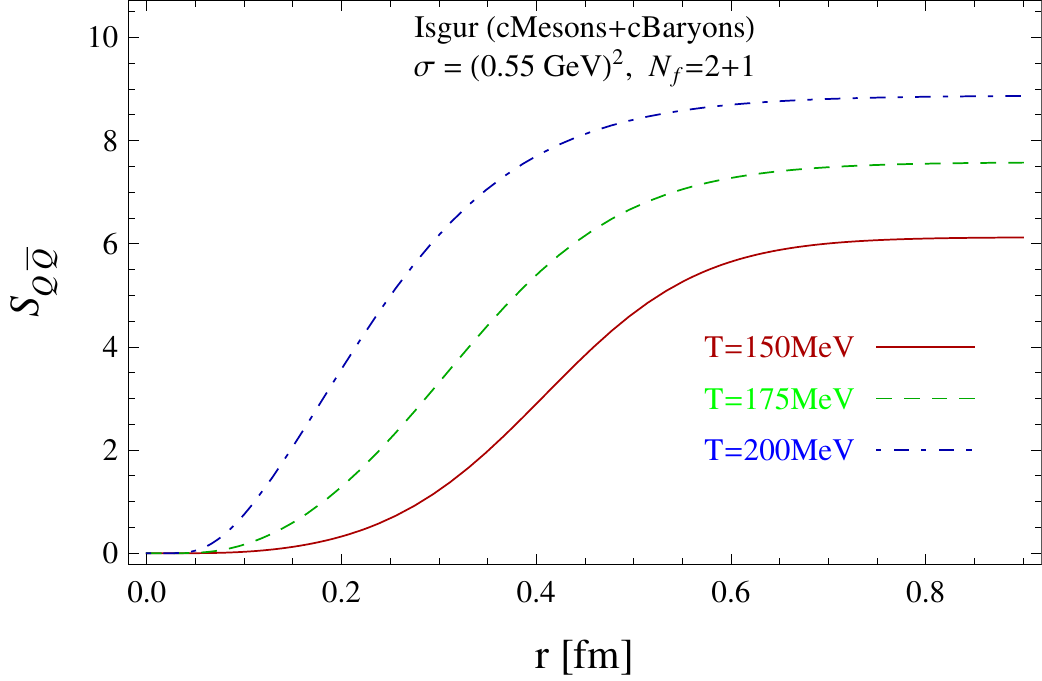,height=60mm,width=85mm}
\epsfig{figure=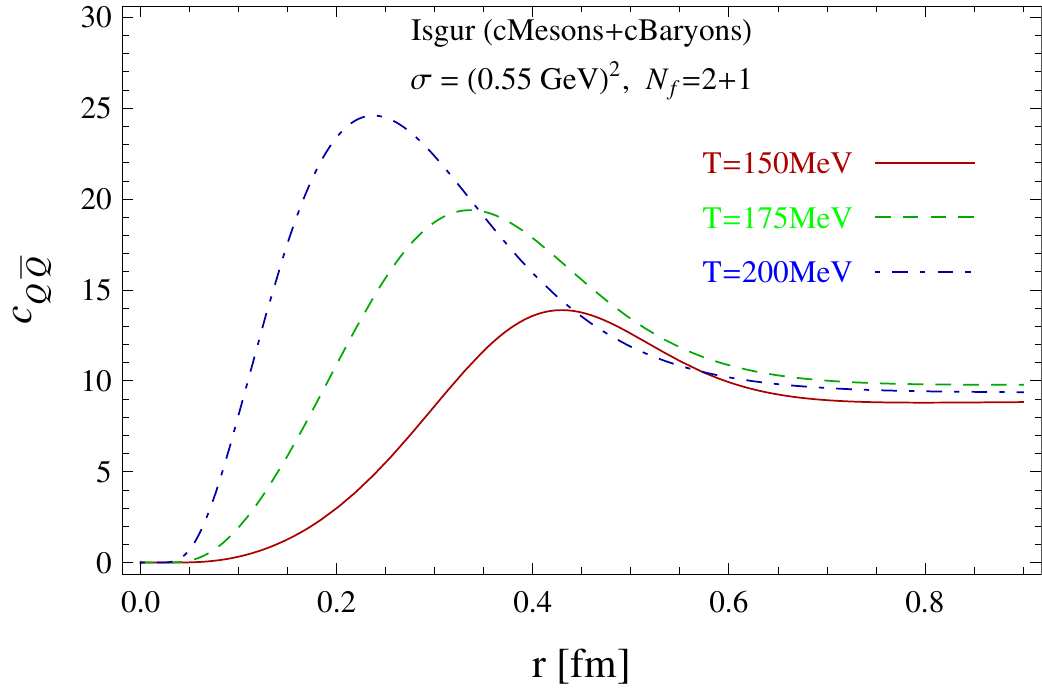,height=60mm,width=85mm}
\end{center}
\caption{With the same model as in Fig.~\ref{fig:Vqq}, value of the entropy
  (upper panel) and specific heat (lower panel) of two heavy sources as a
  function of separation with the model in the absence of mixing given by
  Eq.~(\ref{eq:Fave2}). We consider the spectrum of heavy-light mesons and
  baryons with a charm quark, obtained with the Isgur model of
  Ref.~\cite{Godfrey:1985xj}.}
\label{fig:plotS}
\end{figure}

\begin{figure}[t]
\begin{center}
\epsfig{figure=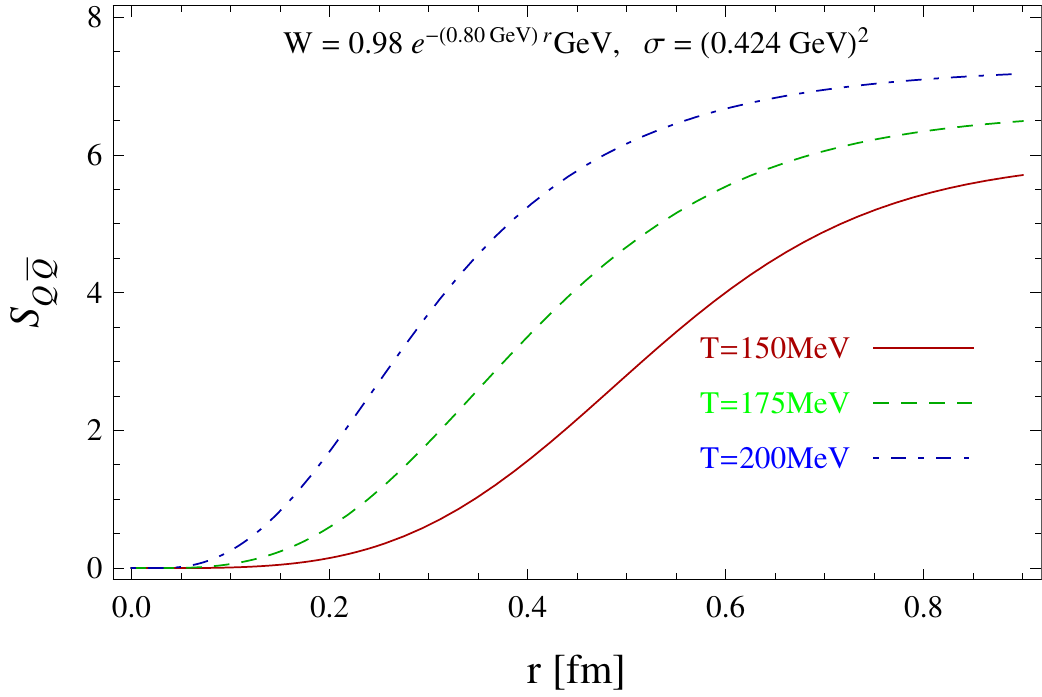,height=60mm,width=85mm}
\epsfig{figure=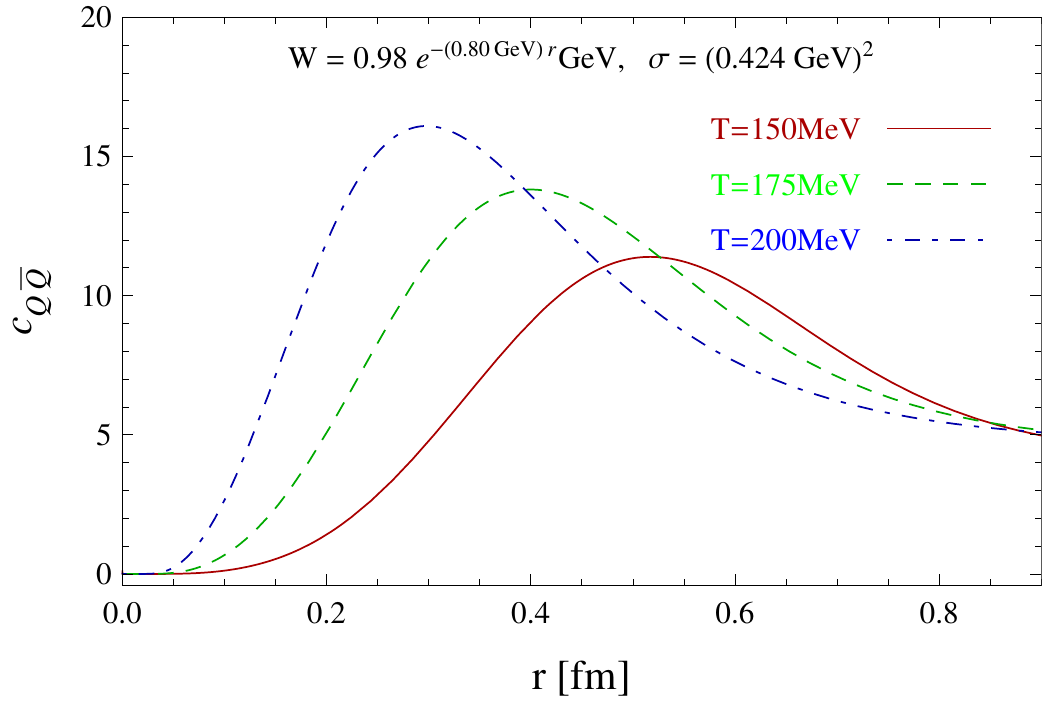,height=60mm,width=85mm}
\end{center}
\caption{With the same model of Fig.~\ref{fig:plotW} (up), value of the entropy (upper panel) and specific heat (lower panel) of
  two heavy sources as a function of separation with the two states model
  of~\Eq{V2states}.}
\label{fig:plotcv}
\end{figure}

An obvious application of the HRG is the determination of the entropy and the
specific heat shifts due to the addition of the extra heavy sources located at
a separation distance $r$, in the confined phase.  On the lattice, such a
calculation was carried out in Ref.~\cite{Kaczmarek:1900zz}, just for the
entropy, using the short distance renormalization of the free energy, which we
have already analyzed above. There one observes negative entropy (shifts)
above the phase transition.  Unfortunately, the lattice calculation of
Ref.~\cite{Borsanyi:2015yka}, which implements a finite distance
renormalization condition, does not provide the entropy; for the available
$\Delta T = 25\MeV$, a sizable noise in the numerical extraction of the
derivative is expected.

In Figs.~\ref{fig:plotS} and \ref{fig:plotcv} we show our model results for the
entropy and specific heat shifts in two different cases. In
Fig.~\ref{fig:plotS} we display the results in the absence of mixing,
Eq.~(\ref{eq:Fave2}), whereas in Fig. \ref{fig:plotcv} we include the avoided
crossing effects as discussed in Sec.~\ref{subsec:avoided_crossings}. Note
that both quantities vanish in the limit $r \to 0$. This was expected as in
this limit the heavy quark and antiquarks are placed at the same point, so
there are no effects of polarization on the thermal medium. At large distances
both quantities tend to stabilize. In effect, in this limit $F_{Q\bar{Q}}(r,
T) \to 2 F_Q(T)$ which is independent of $r$, and the same applies for any
derivatives of the free energy. We remind that our model calculation assumes
that the entropy shift can be treated as a true entropy; this is a valid
assumption at low temperatures, as discussed repeatedly in this work. With
this proviso in mind, we note that both quantities are positive. However,
while this entropy is an increasing function with $r$, this is not always true
for the specific heat. The existence of a peak in $c_{Q\bar{Q}}$ at some
distance $r=r_*$ might be related to the distance at which string breaking
effects take place, being this distance smaller for higher temperatures.

\section{Conclusions}
\label{sec:Conclusions}

In the present paper we have shown that below the phase transition the
partition function of the thermal QCD medium in presence of two
conjugated heavy charges can be expressed in terms of pairs of purely hadronic
states with a single heavy-quark plus the quark-antiquark static potential of
gluodynamics and actually this description provides a gateway to determine the
string tension. This implies a non-trivial level avoided crossing structure in
the heavy-light hadronic spectrum which successfully describes lattice
QCD data.  This is done in harmony with a new K\"all\'en-Lehmann spectral
representation of the Polyakov correlator, whence concavity properties of the
free-energy at any temperature can be straightforwardly deduced.  In addition,
since quite generally any quantum system undergoing a conventional thermal
phase transition will be unbounded in some direction, our construction (in
which $1/T$ is a spatial size of the rotated system) suggests that {\em
  thermal} phase transitions can be regarded as {\em quantum} phase
transitions at zero temperature, the external parameter being the physical
temperature $T$. As it is well known the order of the quantum phase transition
is related to the level structure and corresponding avoided crossing pattern
in the $\mu_n (T)$ eigenvalues~\cite{carr2010understanding}. As an exciting
speculation motivated by our exact spectral representation, one may wonder
whether the string tension admits an exact definition (complex and perhaps
$T$-dependent) even in unquenched QCD, in the same way that the masses of
hadronic resonances admit a precise definition as poles of the $S$-matrix
complex plane. The Dirac-delta term in \Eq{7} remains an exact contribution in
gluodynamics, and this term might move to the complex $\mu$-plane in full QCD.
Finally, a direct determination of the entropy and specific heat shifts due to
separated colour charge conjugated sources from the string-hadron model has
been undertaken with the hope to motivate accurate lattice results of both
quantities.

\begin{acknowledgments}

This work was supported by Spanish Ministerio de Econom\'{\i}a y
Competitividad and European FEDER funds under contract FIS2014-59386-P, by
Junta de Andaluc\'{\i}a grant FQM-225. The research of E.M. is supported by
the European Union under a Marie Curie Intra-European Fellowship
(FP7-PEOPLE-2013-IEF) with project number PIEF-GA-2013-623006.
\end{acknowledgments}

\appendix
\section{Derivation of the K\"all\'en-Lehmann spectral representation}
\label{app:AKL}

In this appendix we present the derivation of the K\"all\'en-Lehmann spectral
representation for the correlation function of the two conjugated Polyakov
loops separated a distance $r$. Let us set one Polyakov loop at $\bm{0}=(0,0,0)$
and the other at $\vx=(x,y,z)$,
\begin{equation}
C(\vx) = \langle 0,T| \mathcal{T}
(\tr_R\Omega(\vx) \tr_R \Omega^\dagger(\bm{0}))|0,T\rangle
.
\end{equation}
The chronological ordering operator $\mathcal{T}$ indicates that the order of
the product must be reversed if $z<0$. Such ordering is automatically
implemented in the functional formulation. Three dimensional rotational
invariance allows to use $|z|$ instead of $z$.

Since the Hamiltonian $H_z(T)$ implements the evolution with respect to the
coordinate $z$ (regarded as an imaginary time) and $\vP_\perp = (P_x,P_y)$
implements the spatial translations on the plane $(x,y)$, on can write
\begin{equation}
C(\vx) = \langle 0,T| 
\tr_R\Omega e^{i \vx_\perp \vP_\perp }  e^{- |z| H_z(T)} \tr_R \Omega^\dagger|0,T\rangle
.
\end{equation}
Inserting a complete set of eigenstates of $\vP_\perp$ and $H_z(T)$,
\begin{equation}
C_c(\vx) = \sum_{n > 0} 
 e^{i \vx_\perp \vp_{\perp,n} }  e^{-|z| w_n(T)} 
|\langle n,T|  \tr_R \Omega^\dagger|0,T\rangle|^2
,
\end{equation}
where we have already removed the disconnected vacuum contribution and
$w_n(T)>0$. The result can be expressed as
\begin{equation}
C_c(\vx) = 
\int \frac{d^2p_\perp}{(2\pi)^2}\frac{dw}{2\pi} e^{-|z|w+i\vx_\perp \vp_\perp}
\rho(w,\vp_\perp)
,
\label{eq:A1}
\end{equation}
where
\begin{equation} \begin{split}
\rho(w,\vp_\perp) &=
\sum_{n > 0} (2\pi)^2\delta(\vp_\perp-\vp_{\perp,n}) 2\pi \delta(w-w_n(T))
\\
& \times
|\langle n,T|  \tr_R \Omega^\dagger|0,T\rangle|^2
.
\end{split}\end{equation}
Now we can invoke three dimensional Lorentz invariance with respect to
$(x,y,-iz)$, or equivalently, three dimensional rotational invariance with
respect to $(x,y,z)$, to conclude that $\rho(w,\vp_\perp)$ is really a
function of $w^2-\vp_\perp^2$. Thus
\begin{equation}
\rho(w,\vp_\perp) = \theta(w) \int_0^\infty d\mu \tau(\mu) 2\pi
\delta(w^2-\vp_\perp^2-\mu^2),
\end{equation}
where the spectral mass density function $\tau(\mu)$ is non negative.

Inserting this form of $\rho$ in \Eq{A1}, and using the relations
\begin{equation}\begin{split}
&
\int_0^\infty \frac{dw}{2\pi}e^{-|z|w} 2\pi \delta(w^2 - w_0^2)
=
\frac{e^{-|z|w_0}}{2w_0}
\\ &
=
\int_{-\infty}^{+\infty} \frac{dp_z}{2\pi} \frac{ e^{i z p_z} }{p_z^2 + w_0^2
  }
,\quad
w_0>0
\end{split}\end{equation}
gives (for $w_0=\sqrt{p_\perp^2 + \mu^2}$)
\begin{equation}\begin{split}
C_c(\vx) &= 
\int_0^\infty d\mu \tau(\mu) 
\int \frac{d^3p}{(2\pi)^3} 
\frac{e^{i\vx \vp}}{p^2+\mu^2}
\\&=
\int_0^\infty d\mu \tau(\mu) \frac{e^{-\mu r}}{ 4\pi r}
.
\end{split}\end{equation}
Explicit three dimensional rotational invariance is recovered in the final
result.

\vskip 50mm


\end{document}